\documentclass[%
 reprint,
 amsmath,amssymb,
 aps, prx,
]{revtex4-2}

\usepackage{graphicx}
\usepackage{dcolumn}
\usepackage{bm}
\usepackage{float}

\usepackage{algpseudocode}
\usepackage{algorithm}
\usepackage{physics}
\usepackage{color}

\usepackage[colorlinks=true, linkcolor=blue, citecolor=blue, urlcolor=blue]{hyperref}
\usepackage[capitalise]{cleveref}
\usepackage[all]{hypcap}

\usepackage{amsmath} 						
\usepackage{amssymb} 						
\usepackage{mathtools}			
\usepackage{mathrsfs}                       
\usepackage{amsthm}
\usepackage{amssymb}
\usepackage{hyperref}
\hypersetup{
    colorlinks = true,
    linkcolor = blue,
    allcolors=blue
    }

\newtheorem{theorem}{Theorem}

\newtheorem{definition}{Definition}

\usepackage{algpseudocode}
\usepackage{subcaption}

\begin{document}

\preprint{APS/123-QED}

\title{From Krylov Complexity to Observability: Capturing Phase Space Dimension with Applications in Quantum Reservoir Computing}


\author{Saud \v{C}indrak}
\email{saud.cindrak@tu-ilmenau.de}
\affiliation{%
 Technische Universität
Ilmenau, Institute of Physics, Ilmenau, Germany
}%

\author{Kathy L{\"u}dge}
 \affiliation{%
 Technische Universität
Ilmenau, Institute of Physics, Ilmenau, Germany
}%
\author{Lina Jaurigue}
\affiliation{%
 Technische Universität
Ilmenau, Institute of Physics, Ilmenau, Germany
}%

\date{\today}

\begin{abstract}
We demonstrate that time-evolved operators can construct a Krylov space to compute Operator complexity and introduce Krylov observability as a measure of effective phase space dimension in quantum systems. We test Krylov observability in the framework of quantum reservoir computing and show that it closely mirrors information processing capacity, a data-driven expressivity metric, while achieving computation times that are orders of magnitude faster. Our results validate Operator complexity and give the interpretation that data in a quantum reservoir is mapped onto the Krylov space. 
\end{abstract}

\maketitle

\textit{Introduction.} --- The significance of physics in machine learning, and vice versa, is unmistakable, as evidenced by the awarding of the Nobel Prize to John Hopfield and Geoffrey Hinton. 
An example of the intersection of physics and machine learning is quantum computing, a technology that seeks to harness quantum effects for computation. 
Key research questions in quantum computation are the extent to which machine learning algorithms can be implemented on quantum systems, how they might function, and what advantages this could yield. Therefore, understanding quantum machine learning networks is of utmost importance. In classical machine learning, understanding is achieved through methods of expressivity and explainability. In quantum mechanics, quantum information measures are used to gain deeper insights. This raises the question of how advancements in quantum information can be leveraged to enhance our understanding of quantum machine learning.

A promising information measure in this context is Krylov complexity, which quantifies how much the evolved operators or states spread within the Krylov basis \cite
{PAR19, BAL22}. It has been shown that Operator complexity measures provide insights into quantum chaos \cite{BAL23a}, spin chains, SYK models \cite{BAL22,BHA24a, JIA21}, 2D CFT \cite{CAP21}, closed systems, driven systems \cite{NIZ23}, and open quantum systems \cite{BHA22a, LIU23b}, among many others \cite{AGU24, ALI23, ANE24, BAL22, BAR19, CAM23, CAP24, CAO21, CHA23, CRA23, DYM21, ERD23, FAN22, GAU23, GIL23, GUO22A, HAS23, HE22, HEV22, HUH24, IIZ23, KIM22, LI24, LIN22, MAG20, MUC22, NAN24, NIE06a, PAL23, PAT22, RAB23, VAS24, CIN24a, ZHO25b, NAN24a, SUC25, BHA22b,BHA22}. Operator complexity can be interpreted as the expectation value of the associated Krylov operator. Although operator and spread complexity have proven to be successful measures to further our understanding, they provide only limited insight into the functioning of a quantum machine learning network \cite{CIN25a}. 

\begin{figure}[t]
\centering
    \hspace*{-0.5 cm}
    \centering
    \includegraphics[scale=1]{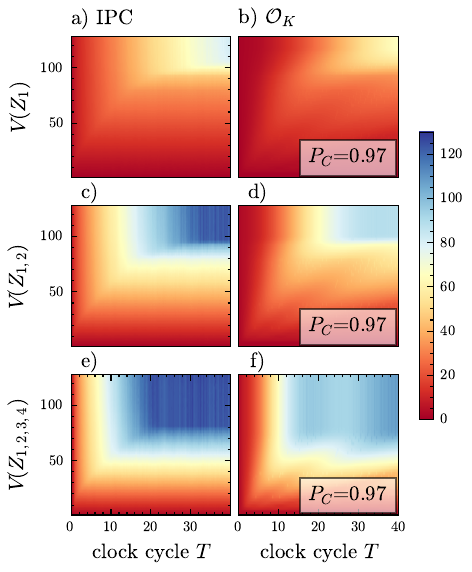}
    \caption[Observability]{ $\mathrm{IPC}$ (a,c,d) and Krylov observability $\mathcal{O}_K$ (b,d,f) in dependence on multiplexing $V$ and clock cycle $T$ when measuring one (a,b), two (c,d) and four (e,f) sites.}
    \label{fig:observability}
\end{figure}

To address the problems with Krylov complexity, we propose \textit{Krylov observability} as a measure to effectively capture the phase space dimension. In contrast to operator complexity, Krylov observability can accommodate multiple operators, respects an upper bound defined by the effective phase space dimension, and integrates multiple measurements, which are all necessities for quantum machine learning. The proposed techniques can be extended to define a measure of operator complexity that effectively accounts for multiple operators. Interestingly, a similar approach has been made in \cite{CRA25}, where authors propose multiseed operator complexity as a robust measure.
We test these measures in the context of quantum reservoir computing. Unlike variational quantum machine learning, quantum reservoir computing utilizes a physical system in which only the readout layer is trained \cite{JAE01, MAA02, VER07, APP11, FUJ17}. Extensive research has been conducted to optimize quantum reservoir computing, ranging from leveraging noise and decoherence, improving task performance and developing various methods to address the time-complexity problem \cite{FUJ17, NAK19, KUT20, MAR21, XIA22, GOE23, CHE20b, SUZ22, PFE22, YAS23, NOK24, ABB24a, SAN24, FRY23, DOM23, MUJ23, CIN24, AHM24, KOB24, ZHU24a, SCH25, VET25, SCH25}. An established measure of expressivity in reservoir computing is the information processing capacity ($\mathrm{IPC}$) \cite{DAM12}, which quantifies the reservoir's ability to map and retain data both linearly and non-linearly. 
In this work we show that time-evolved operators can construct a Krylov space to compute Operator complexity and introduce Krylov observability as a measure of the effective phase space dimension. We then compute Krylov observability for various systems and observables, and compare the behavior with the information processing capacity. We demonstrate that Krylov observability and information processing capacity exhibit correlation factors as high as 0.97. Our results show that quantum reservoirs map data onto the Krylov space and that sampling this space can be indicative of task performance. Our findings demonstrate the utility of Krylov observability as a measure of the computational capacity of quantum systems which is both upper bounded and computationally inexpensive. Calculating the information processing capacity for the results in Figure \ref{fig:observability} took 150 hours per reservoir, whereas Krylov observability was computed in just 30 seconds per reservoir. Lastly, we introduce a characteristic timescale for operators, motivated by the quantum Zeno effect \cite{MIS77}, which effectively describes the rate at which IPC and Krylov observability increase, and takes a form similar to the quantum Zeno time discussed in \cite{FAC08}. 

Spread complexity was first discussed in \cite{DOM24} in relation to quantum reservoir computing, where the authors show that spread complexity can be used to explain task performance.
In \cite{CIN25a}, we perform a thorough study of various Krylov-based measures in relation to understanding task performance in QRC. We discuss Krylov observability, which is introduced in this work, as well as fidelity, spread complexity \cite{BAL22}, Krylov expressivity \cite{CIN24a}, and the information processing capacity in the context as expressivity measures for quantum reservoir computers. There, we show that more expressive quantum systems can exhibit worse task performance if the considered observables commute with a larger subspace of the system.

\textit{Krylov spaces} --- The evolution of an operator $O$ with a Hamiltonian $H$ is given by $\partial_t O(t) = i[H, O(t)]$.
The solution to this equation is
\begin{align}
    O(t) &= e^{iHt}Oe^{-iHt} = \sum_{k=0}^\infty \frac{(it)^k}{k!}\mathcal{L}^k(O), \nonumber
\end{align}
where $\mathcal{L}(O) = HO - OH$ represents the Liouvillian. This implies that $O(t) \in \mathrm{Span}\{ \mathcal{L}^0(O), \mathcal{L}^1(O),\ldots \}$. The authors in \cite{PAR19} used the Krylov space property and the linearity of $\mathcal{L}(O)$ to show that there exists an $M \in \mathbb{N}$ such that 
\begin{align}
\label{eq:KrylovL}
    O(t) \in \mathrm{Span}\{\mathcal{L}^0(O), \mathcal{L}^1(O), \dots, \mathcal{L}^{M-1}(O)\} = \mathcal{L}_M 
\end{align}
holds true. Orthonormalizing the Krylov space $\mathcal{L}_M$ results in the basis $\{\mathcal{W}_i\}_{i}$, such that the space can be rewritten as $\mathcal{L}_M = \mathrm{Span}\{\mathcal{W}_0, \mathcal{W}_1, \dots, \mathcal{W}_{M-1}\}$. With the representation of the time-evolved operator $O(t) $ in the Krylov basis, the Operator complexity $ \mathcal{K}_O(t) $ is defined as follows \cite{PAR19}:
\begin{align}
    O(t) &= \sum_{n=0}^{M-1} i^n \beta_n(t) \mathcal{W}_n,~~~
   \mathcal{K}_O(t) &= \sum_{n=0}^{M-1} (n+1) \abs{\beta_n(t)}^2.
    \label{eq:spread_complexity}
\end{align}
\textit{Krylov Observability}---
In this work, we will prove that time-evolved observables can be used to construct a space $\mathcal{F}_M$ that is equivalent to the space $\mathcal{L}_M$. Building upon these results, we will define an observability measure to predict task performance and further our understanding of quantum reservoir computing.

\begin{theorem}[]
\label{theorem:timeevolv}
    Let $O \in \mathbb{C}^{N,N}$ be a Hermitian observable, i.e., $O = O^\dagger$. Let $O(t) = e^{iHt}Oe^{-iHt}$, where $H$ is the Hamiltonian, and let $M$ be the Krylov grade, such that $O(t) \in \mathcal{L}_M$. Let $0 = t_0 < t_1 < t_2 < \dots < t_Q$ and define the approximate evolved operator as
    \begin{align*}
        \Tilde{O}(t_a) = \sum_{k=0}^Q \frac{(it_a)^k}{k!} \mathcal{L}^k(O).
    \end{align*}
    Then, it holds that $\mathcal{L}_Q = \Tilde{\mathcal{F}}_Q$, for all $Q \in \mathbb{N}$, where $\Tilde{\mathcal{F}}_Q$ is given by
    $
    \Tilde{\mathcal{F}}_Q = \mathrm{Span}\{\Tilde{O}(t_0), \Tilde{O}(t_1), \dots, \Tilde{O}(t_Q)\}$.
    In the limit as $Q \rightarrow \infty$, it follows that $\mathcal{L}_M = \mathcal{F}_M$, where $\mathcal{F}_Q$ is given by 
    $
    {\mathcal{F}}_Q = \mathrm{Span}\{{O}(t_0), {O}(t_1), \dots, {O}(t_{M-1})\}
    $.
    \begin{proof}
    $\Tilde{O}(t_a)$ can be represented as 
    \begin{align*}
        \Tilde{O}(t_a) &= (i^0 \mathcal{L}^0(O), i^1 \mathcal{L}^1(O), \dots, i^Q \mathcal{L}^Q(O)) 
        \begin{pmatrix}
            1\\
            t_a\\
            \vdots\\
            \frac{t_a^Q}{Q!}
        \end{pmatrix}
    \end{align*}
    Representing $Q + 1$ time-evolved $\Tilde{O}(t_a)$ with pairwise distinct times $t_0 < t_1 < \dots < t_Q$ leads to
    \begin{align*}
        &(\Tilde{O}(t_0), \dots, \Tilde{O}(t_Q))= (i^0 \mathcal{L}^0(O), \dots, i^Q \mathcal{L}^Q(O)) \Theta, \\
        \mathrm{with~}&\Theta =   \begin{pmatrix}
            1 & 1 & 1 & \dots & 1 \\
            {t_0}/{1!} & {t_1}/{1!} & {t_2}/{1!} & \dots & {t_{Q}}/{1!} \\
            {t_0^2}/{2!} &{t_1^2}/{2!} & {t_2^2}/{2!} & \dots & {t_{Q}^2}/{2!} \\
            \vdots & \vdots & \vdots & \vdots & \vdots \\
            {t_0^{Q}}/{Q!} & {t_1^{Q}}/{Q!} & {t_2^{Q}}/{Q!} & \dots & {t_{Q}^{Q}}/{Q!}
        \end{pmatrix}.
    \end{align*}
    The pairwise distinct times $t_a$ imply the linear independence of each column and the invertibility of $\Theta$, resulting in $$(\Tilde{O}(t_0), \dots, \Tilde{O}(t_Q)) \Theta^{-1} = (i^0 \mathcal{L}^0(O), \dots, i^Q \mathcal{L}^Q(O)).$$
    Since all vectors of $\mathcal{L}_Q$ (see \cref{eq:KrylovL}) can be represented through vectors in $\Tilde{\mathcal{F}}_Q$, it follows that $\mathcal{L}_Q =\Tilde{\mathcal{F}}_Q$ holds for all $Q$. In the limit as  $Q$ approaches infinity, $ \tilde{O}(t_a) \rightarrow O(t_a) $ holds, such that the space $ \tilde{\mathcal{F}}_Q $ can be expressed as:
    \begin{align}
       \mathcal{L}_Q=\Tilde{\mathcal{F}}_Q = \mathrm{Span}(O(t_0), O(t_1), \dots, O(t_Q))={\mathcal{F}}_Q.
    \end{align}
    For $Q > M$, $\mathcal{L}_Q$ has only $M$ linearly independent vectors, implying that $\mathcal{F}_Q$ has only $M$ linearly independent vectors. We have thus shown that the Krylov space generated by the powers of the Liouvillian $\mathcal{L}^k$ (\cref{eq:KrylovL}) is equivalent to the Krylov space with $M$ time-evolved observables $O(t_k)$ with pairwise different times, i.e. $\mathcal{L}_M = \mathcal{F}_M$.
    \end{proof}
\end{theorem}
Up to this point, we have shown that there are $M$ times, so that $\mathcal{L}_M = \mathrm{Span}\{O(t_1),\ldots,O(t_M) \}$ applies, but not what these times $t_j$ are. For the case of equidistant sampling times $t_j=(j-1)\tau$ we can define the linear function $g(O)=UOU^\dagger$. The space of $Q$ time-evolved observables results in $\mathcal{F}_Q=\mathrm{Span}\{O,g(O), g^2(O),\ldots, g^Q(O)\}$ being a Krylon space. Since $\mathrm{dim}(\mathcal{F}_Q) = M$ and $\mathcal{F}_Q$ is a Krylov space, it follows from the properties of Krylov spaces that the first $M$ time-evolved observables are linearly independent. This implies that $\mathcal{L}_M = \mathcal{F}_M = \mathrm{Span}(O, g(O), g^2(O), \dots, g^{M-1}(O))$, and that the first $M$ times, $t_j \in \{0, \tau, 2\tau, \dots, (M-1)\tau\}$, can be used.

In QRC, the expectation values of \( K \) observables \( O_1, O_2, \dots, O_K \) are sampled at discrete times \( \ev{O_i(t_j)} \). From \cref{theorem:timeevolv}
\begin{align}
     O_i(t) \in \mathcal{F}_i = \mathrm{Span}(O_i(t_0), O_i(t_1), \dots, O_i(t_{M_{i-1}}))
\end{align}
follows, where $M_i$ is the dimension of $\mathcal{F}_i$, and $ t_{M_{i-1}} = T_O $ is equal for all observables $ O_i $.
For any $ O_i \in \{O_1,O_2,\ldots,O_K\} $, it holds that
\begin{align}
    O_i(t) \in \bigcup_{i=1}^K \mathcal{F}_i =: \mathcal{F}(O_1, O_2, \dots, O_K), \quad \forall O_i.
\end{align}
In the following, \( \mathcal{F} \) always denotes the full space, i.e., \( \mathcal{F} := \mathcal{F}(O_1, O_2, \dots, O_K) \) and since each time-evolved observable is within this space, we address this as phase space. Constructing the spaces in this way does not account for the linear independence of the individual spaces \( \mathcal{F}_k \) and does not provide information about how much each \( \mathcal{F}_k \) contributes to \( \mathcal{F} \). 
To address these issues, we construct the spaces according to \cref{alg:cap}, as described in \cref{app:algorithm}. The resulting spaces \( \mathcal{F}_k \) are then linearly independent, ensuring that each observable \( O_k \) is mapped onto a distinct space \( \mathcal{F}_k \). 
In the next step, we use these spaces to define our main result, \textit{Krylov observability}, which serves as a measure to capture the effective phase space dimension of $\mathcal{F} $ associated with multiple operators measured at distinct times. 

The construction of the phas space domension is by constructing  $\mathcal{F}$ from time-evolved operators at times $t_n = nT/M$, where $M$ is the Krylov grade, resulting in a time-dependent $\mathcal{F}(T)$. For each $\mathcal{F}(T)$, we then define a measure of dimensionality, which we call the *effective phase space dimension*. Thus, in contrast to Krylov complexity—where complexity is quantified by how intricate the time evolution is within the Krylov space—here we define a Krylov space for each time $T$.

The phase space dimension is constructed by forming $\mathcal{F}(T)$ from time-evolved operators at times $t_n = nT/M$, where $M$ is the Krylov grade. For each $\mathcal{F}(T)$, we define a measure of dimensionality, referred to as the *effective phase space dimension*. In contrast to Krylov complexity, where complexity is quantified by how intricate the time evolution is within a fixed Krylov space, here a distinct Krylov space is defined for each time $T$.

\begin{definition}[Krylov Observability] 
Assume \( K \) observables \( O_1, \dots, O_K \). For each observable \( O_k \), the spaces \( {\mathcal{F}}_k \) are computed with \( \mathrm{dim}({\mathcal{F}}_k) = M_k \) and let \( T > 0 \). With \( R_k = \mathrm{min}(V, M_k) \) and \( \tau_{k} = T / R_k \), define
$$\mathcal{G}_k = \{O_k(\tau_1), O_k(\tau_2), \dots, O_k(\tau_{R_k})\}$$
where $V$ is the number of measurements for each observable. Upper bounding the number of times $R_k$ by $M_k$ ensures that $\mathcal{G}_k$ always consists of the smallest number of linearly independent basis states.   The observability of the \( k \)-th observable \( O_k \) is defined as 
\begin{align}
    p_k(T) = 1 + \sum_{j=1}^{R_k - 1} (1 - f(O_k(\tau_j), O_k(\tau_{j+1}))),
\end{align}
where \( f \) is a function with \( f(A,B) = 1 \) if \( A = B \). In this study, we set 
\begin{align}
     f(A,B) := \mathrm{F}(A,B) = \abs{\mathrm{Tr}\Big(\frac{A^\dagger B}{\norm{A}\norm{B}}\Big)}
     \label{eq:fidelity}
\end{align}
\( \mathrm{F} \) is the normalized fidelity between the two time-evolved operators. The $\textbf{Krylov ~Observability}$ \( \mathcal{O}_K(T) \) of \( V \) multiplexed observables \( O_1, \dots, O_K \) is defined as  
\begin{align}
    \mathcal{O}_K(T) = \sum_{k=1}^K p_k(T).
\end{align}
\label{def:observability}
\end{definition}

\textit{Quantum Reservoir Computing}---
We demonstrate the utility of Krylov Observability using the widely researched Ising Hamiltonian
\begin{align}
H_{I} = \sum_{i=1,j>i}^{N_S}J_{ij}X_iX_j+\sum_{i=1}^{N_S}hZ_i.
\end{align}
as a quantum reservoir. $X_i$ and $Z_i$ represent the Pauli-$x$ and Pauli-$z$ operators acting on the $i$-th site, and $N_S$ is the number of sites. The constant $h$ is set to $h=0.5$, and $J_{i,j}$ is sampled from a uniform distribution on the interval $[0.25,0.75]$ \cite{FUJ17}. The system evolution under the von Neumann equation is given by
\begin{align}
    \partial_t \rho = -i[H,\rho].
\end{align}

In the quantum reservoir scheme, the $n$-th input of a time series $\mathbf{u} = (u_1, u_2, \ldots, u_{N_{U}})$ is encoded into one site as $\ket{\Psi_n} = \sqrt{\frac{1 - u_n}{2}} \ket{0} + \sqrt{\frac{1 + u_n}{2}} \ket{1}$. Then, the full density matrix is constructed as $\rho_n = \ket{\Psi_n}\bra{\Psi_n} \otimes \Tr_1(\rho_{n-1}(T))$, where $T$, also called the clock cycle, is the evolution time before the next input is introduced. The system evolution for any time $t$ is described by $\rho_n(t) = e^{-iHt} \rho_{n} e^{iHt}$. The output is constructed by computing the expectation values $\ev{O_{n,i}(\tau_j)} = \mathrm{Tr}(O_i \rho_n(\tau_j))$ of $K$ observables $O_1, O_2, \ldots, O_K$ at discrete times $\tau_i = iT/V$. This results in $N_R = VK$ outputs for each input. Writing the expectation values for each input in a row forms the state matrix $\mathbf{S} \in \mathbb{R}^{N_{U}, N_R}$. Gaussian noise $\mathcal{N}$ is added to account for shot noise and serves as regularization, i.e., $\mathbf{S} \leftarrow \mathbf{S} + \eta \mathcal{N}$. The output of the reservoir $\mathbf{Y}$ is given by $\mathbf{Y} = \mathbf{S} \mathbf{W}$, where $\mathbf{W}$ are weights computed on a training set $\mathbf{u}_{\mathrm{Tr}}$ with $\mathbf{W} = (\mathbf{S}_{\mathrm{Tr}}^T \mathbf{S}_{\mathrm{Tr}})^{-1} \mathbf{S}_{\mathrm{Tr}}^T \mathbf{\hat{Y}}_{\mathrm{Tr}}$. \\
The \textbf{Information Processing Capacity (IPC)} is a measure of how effectively a reservoir computing system can reconstruct or generalize prior input data using a set of orthogonal target functions, typically Legendre polynomials \( l_n \) \cite{DAM12}.
We define the \( k \)th polynomial of the input \( \mathbf{u} \) \( d \) steps into the past as 
\(
\mathbf{l}_k(-d) = (l_k(u_{-d}), l_k(u_{-d+1}), \ldots, l_k(u_{N-d})),
\)
where \( N \) is the length of the input series sampled from a uniform distribution, i.e., \( u \in \mathcal{U}([-1,1]) \) \cite{KUB21}.
The first-order IPC, denoted as \( \mathrm{IPC}_1 \), is also referred to as memory capacity or linear short-term memory and is given by
$
\mathrm{IPC}_1 = \sum_d C(\mathbf{Y}, \mathbf{l}_1(-d)).
$
The second-order IPC, \( \mathrm{IPC}_2 \), consists of two terms: one involving the second-order Legendre polynomial \( \mathbf{l}_2(-d) \), and another involving pairwise products of delayed first-order terms, i.e., \( \mathbf{l}_1(-d_1) \mathbf{l}_1(-d_2) \). This can be extended to higher-order IPCs, denoted \( \mathrm{IPC}_n \), and summing over all orders results in the total IPC, or simply IPC, given by
$
\mathrm{IPC} = \sum_i \mathrm{IPC}_i.
$
The IPC is therefore a hypertask, requiring up to thousands of training runs—one for each combination of delays \( d \). A detailed description of IPC and a discussion of the memory–nonlinearity trade-off in quantum reservoir computing (QRC) is provided in \cite{CIN24}. 

\textit{Results}---To analyze the ability of our Krylov observability measure $\mathcal{O}_K$ to quantify the computational capabilities of a quantum system, we compare it with the 
$\mathrm{IPC}$. We consider ten different Ising Hamiltonians, each with four sites and uniformly distributed coupling constants $J_{i,j}$ on the interval $[0.25, 0.75]$. For each Hamiltonian, we consider a  quantum reservoir where the first site ($Z_1$), the first two sites ($Z_{1,2}$), and all four sites ($Z_{1,2,3,4}$) are measured in the Pauli-$z$ direction. We compute the information processing capacity ($\mathrm{IPC}$) and our proposed Krylov observability ($\mathcal{O}_K$) for various clock cycle lengths ($T$) and different time-multiplexing measurements ($V$)

Figure~\ref{fig:observability} shows the results for the averaged $\mathrm{IPC}$ (a,c and d) and the averaged $\mathcal{O}_K$ (b, e, f). In each plot, the $y$-axis represents the number of measurements for the corresponding measurement scheme and the $x$-axis represents the clock cycle $T$. The first row shows results when only the first site is measured $Z_1$. 

\begin{figure}[t]
\centering
    \hspace*{-0.5 cm}
    \centering
    \includegraphics[scale=01]{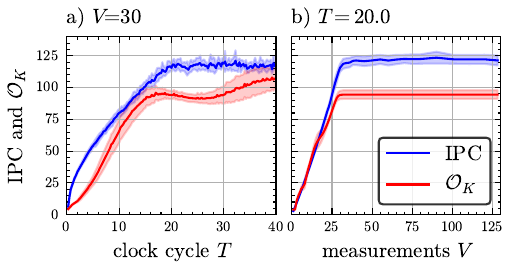}
    \caption[Observability]{Observability $\mathcal{O}_K$ and $\mathrm{IPC}$ in dependence on the clock cycle $T$ in a) and in dependence on multiplexing $V$ in b).}
    \label{fig:1d_curve}
\end{figure}

As the clock cycle \( T \) and the number of measurements \( V \) increase, so does the phase space dimension, i.e., the Krylov observability \( \mathcal{O}_K \), onto which the data is mapped. At \( T_{\mathrm{sat}}(Z_{1}) \approx 36 \) and \( V_{\mathrm{sat}}(Z_{1}) \approx 110 \), the full Krylov space is explored, as indicated by the saturation of \( \mathcal{O}_K \) in \cref{fig:observability}.b). Further increases in the number of measurements ($V$) and clock cycles ($T$) result in measurements that are linearly dependent on all previous ones, thus not further increasing the expressivity of the reservoir. This is observed in the almost identical behavior of \( \mathrm{IPC} \) and \( \mathcal{O}_K \), see \cref{fig:observability}.a) and b).
To better quantify the behavior of the Krylov observability $\mathcal{O}_K$ and information processing capacity $\mathrm{IPC}$, we compute the Pearson correlation factor between them. In the case where only one observable is measured, a Pearson correlation factor of $P_C=0.97$ is reached. Measuring the first two sites yields results in the second row of (\cref{fig:1d_curve}.b and d), where both measures show similar behavior with slightly smaller values for $\mathcal{O}_K$. The number of virtual nodes at which $\mathrm{IPC}$ and $\mathcal{O}_K$ saturate reduces to $T_{\mathrm{sat}}(Z_{1,2})\approx 30$ and $V_{\mathrm{sat}}(Z_{1,2})\approx 95$ with $P_C=0.97$. Measuring all four sites moves the saturation to smaller values, $V_{\mathrm{sat}}(Z_{1,2,3,4})\approx 75$ and $T_{\mathrm{sat}}(Z_{1,2})\approx 18$ (see \cref{fig:observability}.e and f). The earlier saturation when measuring multiple observables is due to the higher number of measurements $N_R$ as the total number of measurements is given by $N_R=VK$.
 Additionally, the space dimension of $\mathcal{F}(Z_1)$ is smaller than that of $\mathcal{F}(Z_{1},Z_{2})$ and $\mathcal{F}(Z_{1},Z_{2},Z_{3},Z_{4})$. One interesting observation, particularly for $\mathcal{O}_K$ in \cref{fig:observability}.f), is oscillations over time, arising from the observability measure construction. To better understand this behavior, \cref{fig:1d_curve} shows the $\mathrm{IPC}$ (blue) and $\mathcal{O}_K$ (red) as functions of the clock cycle \( T \) for a fixed number of measurements \( V = 30 \) (a), and as functions of \( V \) for a fixed clock cycle \( T = 20 \) (b).
In \cref{fig:1d_curve}.a), $\mathcal{O}_K$ increases slightly later but reaches a maximum with oscillations, while the $\mathrm{IPC}$ reaches a maximum around $T\approx 20$ and remains stable. The standard deviation is shown as a fill in the corresponding color and is small for both systems. The oscillation is due to the small number of qubits, as quasiperiodic behavior is more apparent \cite{CIN24a}.
The Krylov observability $\mathcal{O}_K$ and $\mathrm{IPC}$ with $T = 20$, as a function of the number of measurements $V$, show almost identical behavior, except for an offset (see \cref{fig:1d_curve}.b). 

Another important aspect is the rate at which the IPC and Krylov observability increase. To analyze this, we introduce a characteristic time $\tau_z$ for the fidelity defined in \cref{eq:fidelity}, inspired by the quantum Zeno effect \cite{MIS77, FAC08}. This time characterizes the initial linear growth of both IPC and Krylov observability with respect to the clock cycle $T$. 
\begin{figure}[t]
\centering
    \hspace*{-0.5 cm}
    \includegraphics[scale=1]{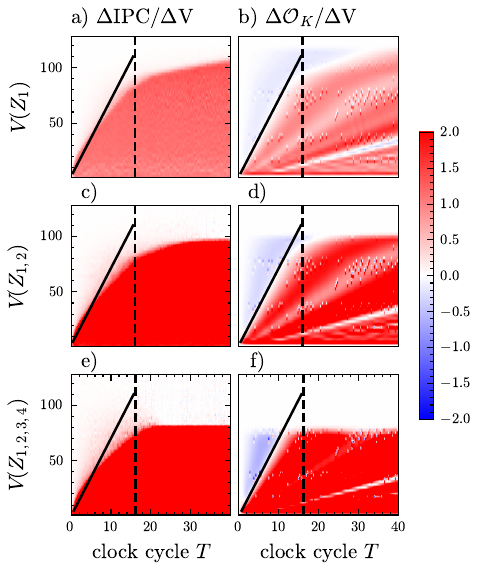}
    \caption[Observability]{
    Change in normalized performance metrics—$\Delta \mathrm{IPC}/\Delta V$ (a, c, e) and $\Delta \mathcal{O}_K/\Delta V$ (b, d, f)—as a function of the clock cycle $T$ and the multiplexing level $V$. The panels show the effect of increasing the number of simultaneously measured sites: one (a, b), two (c, d), and four (e, f). Positive values (red) indicate performance improvement with additional multiplexing, while negative values (blue) indicate degradation. The black dashed and solid lines represent theoretical benchmarks for comparison.
    }
    \label{fig:zeno_time}
\end{figure}
\Cref{fig:zeno_time} shows the change in IPC and $\mathcal{O}_K$ as a function of multiplexing, defined as $\Delta \mathrm{IPC}/\Delta V = \left[\mathrm{IPC}(V_2) - \mathrm{IPC}(V_1)\right]/(V_2 - V_1)$ for neighboring points in the heatmap. We overlay the curve $\tau = \tau_z V$, where $\tau_z$ is the characteristic Zeno time. $\tau_z$ is computed for each observable $Z_1,~Z_2,~Z_3$, and $Z_4$, and averaged over all 10 Hamiltonians used in the statistical analysis. The vertical dashed black line represents the Heisenberg time $t_H$, estimated from the level spacing statistics. Ordering the eigenvalues as $\varepsilon_0 < \varepsilon_1 < \dots < \varepsilon_n$, the level spacings $s_i = \varepsilon_{i+1} - \varepsilon_i$ yield an average spacing $\langle s \rangle$. The Heisenberg time is then given by $t_H = 2\pi / \langle s \rangle$, representing the timescale over which the state explores the Hilbert space \cite{OGA07, BOH84, DEU91, SRE99}. 
The black solid line $\tau = \tau_z V$ closely follows the initial growth in both IPC and $\mathcal{O}_K$ in the top row. Beyond this curve, the decrease in $\Delta \mathcal{O}_K / \Delta V$ is a signature of the quantum Zeno effect: increasing $V$ causes the node separation $T/V$ to fall below $\tau_z$ ($T/V<\tau_z$), thereby reducing Krylov observability. This suppression is also observed when measuring two (middle row) and four (bottom row) sites, reflecting the characteristic freezing behavior of the quantum Zeno effect \cite{MIS77, FAC08}. Since the derivation of $\tau_z$ assumes a single observable the fit is therefor best when only one operator is considered (first row of \cref{fig:zeno_time}). The presence of multiple observables accelerates saturation before reaching $t_H$. Nonetheless, for $T \lesssim 10$, the Zeno scaling effectively captures the initial rate of increase in both IPC and Krylov observability over all realization.,

\textit{Conclusion and Discussion}---
This work showed that, instead of computing different powers of the Liouvillian $[H, O]^k$ for the construction of the Krylov space $\mathcal{L}_M$, time-evolved operators $O(t_k)$ can be used to construct an equivalent Krylov space $\mathcal{F}_M$, i.e. $\mathcal{F}_M=\mathcal{L}_M$ (\cref{theorem:timeevolv}). This enabled our work to propose Krylov observability as a measure to capture the effective phase space dimension of evolved observables (see \cref{def:observability}). Our results show that Krylov observability effectively describes the generalization capabilities of a quantum reservoir, as illustrated by its almost identical behavior to the information processing capacity. The high correlation between the information processing capacity and Krylov observability suggests that quantum reservoir computing can be interpreted as a map of classical data onto Krylov spaces, thus providing an interpretation of quantum reservoir computing.
From another point of view, a quantum reservoir computer can be viewed as an experiment to test Krylov complexity-motivated measures. The question posed is: How well can a measure defined on the Krylov space (Krylov observability) predict a quantum system's ability to remember and non-linearly map macroscopic data (i.e., encoded input series) as measured by the information processing capacity? We conclude, that the almost identical behavior of the proposed Krylov observability and information processing capacity validates the field of Krylov complexity as an effective metric to gain understanding of the time evolution. One significant advantage of Krylov observability over information processing capacity is its four orders of magnitude faster computation time, making it a highly efficient tool for analysis. 
Another important contribution of this letter is the introduction of a timescale inspired by the quantum Zeno effect, which effectively captures the relationship between IPC and Krylov observability as a function of the number of measurements and the clock cycle.
Moreover, Krylov observability holds great potential for adaptation to quantum machine learning applications as a physically inspired measure of quantum expressivity.
Given a parameterized quantum circuit \( U(\Theta_n) \), an observable \( O \), and the definition \( O_n^{(l)} := U(\Theta_n)^l O U^\dagger(\Theta_n)^l \), the corresponding Krylov space can be constructed as
\(
\mathcal{F}_n := \mathrm{Span}\left\{ O, O_n^{(1)}, O_n^{(2)}, \ldots, O_n^{(M-1)} \right\}.
\)
An interesting direction for future work would be to quantify how Krylov observability and Krylov complexity evolve during training, or how the Krylov spaces change between two consecutive training examples, that is, between \( \mathcal{F}_n \) and \( \mathcal{F}_{n+1} \). This analysis could offer a new perspective on scaling system sizes and provide insights into how to initialize the parameters \( \Theta_n \). Additionally, it may reveal meaningful behavior related to increasing circuit depth and contribute to a better understanding of the emergence of barren plateaus.

\appendix

\section{Algorithm for the construction of disjiont Krylov spaces}\label{app:algorithm}
Let \( \mathcal{F}^B \) and \( \mathcal{F}^B_i \) be the bases of the spaces \( \mathcal{F} \) and \( \mathcal{F}_i \), i.e., \( \mathcal{F} = \mathrm{Span}(\mathcal{F}^B) \) and \( \mathcal{F}_i = \mathrm{Span}(\mathcal{F}^B_i) \). 

\begin{algorithm}[H]
\caption{Construction of Observability Spaces}\label{alg:cap}
\begin{algorithmic}[1]
\State \( \mathcal{I}_O = \{1, 2, \dots, K\} \)
\State \( \mathcal{T} = \{t_1, t_2, \dots, t_R\} \)
\State \( \mathcal{F}^{(B)} = \emptyset \)
\State \( \tilde{\mathcal{F}}_1^{(B)}, \tilde{\mathcal{F}}_2^{(B)}, \dots, \tilde{\mathcal{F}}_K^{(B)} = \emptyset, \emptyset, \dots, \emptyset \)
\While {$t_j \in \mathcal{T}$}
    \For{$k \in \mathcal{I}_O $}
        \If {$O_k(t_j) \notin \mathcal{F}$} 
        \State \( \mathcal{F}^{(B)} = \mathcal{F}^{(B)} \cup O_k(t_j) \)
        \State \( \tilde{\mathcal{F}}_k^{(B)} = \tilde{\mathcal{F}}_k^{(B)} \cup O_k(t_j) \)
        \EndIf
    \EndFor
\EndWhile
\State \( \mathcal{F}^{(B)} \leftarrow \mathrm{orthonormalize}(\mathcal{F}^{(B)}) \) 
\end{algorithmic}
\end{algorithm}

In the first two lines, the index set \( \mathcal{I}_O \) for the observables and the set of all times \( \mathcal{T} \) are initialized, where \( R > M \) should hold, and in our case, \( R = \mathrm{dim}(H) \). Lines 3 and 4 initialize all bases to the empty set, and line 5 iterates through all times \( t_j \). The second while loop in line 6 iterates through all observables \( O_k \). If the \( k \)-th time-evolved observable \( O_k(t_j) \) is not in the space \( \mathcal{F} = \mathrm{Span}(\mathcal{F}^B) \), then the time-evolved observable is added to the bases \( \mathcal{F}^{(B)} \) and \( \tilde{\mathcal{F}}_k^{(B)} \) in lines 8 and 9. This can be verified by computing the rank of the basis and checking whether it changes. If the rank does not change, i.e., \( \mathrm{rank}(\mathcal{F}^{(B)}) = \mathrm{rank}(\mathcal{F}^{(B)} \cup O_k(t_j)) \), then it follows that \( O_k(t_i) \in \mathcal{F} \) holds as well. In line 13, the basis \( \mathcal{F}^{(B)} \) is orthonormalized, such that spaces with the following properties are returned:
\begin{align*}
    &\mathcal{F} = \bigcup_{k=1}^K \mathcal{F}_k = \bigcup_{k=1}^K \tilde{\mathcal{F}}_k~,~~  \mathrm{dim} \left( \bigcup_{j=1}^l \tilde{\mathcal{F}}_j \right) = \sum_{j=1}^l \mathrm{dim}(\tilde{\mathcal{F}}_j)
\end{align*}

\section{Quantum Zeno Time}\label{app:quantum_zeno}

Let us consider the fidelity overlap between the observable $O$ and the time-evolved observable at $O(t)$ given by 
\begin{align}
    F(O, O(t))= \abs{\mathrm{Tr}\Big(\frac{O^\dagger O(t)}{\norm{O}\norm{O(t)}}\Big)}
\end{align}
We normalize $O:=O/\norm{O}$. The time-evolved operator up to second order can be approximated as
\begin{align}
    O(t) = e^{iHt} O e^{-iHt}= O + it [H, O] - \frac{t^2}{2} [H, [H, O]] + \mathcal{O}(t^3)
\end{align}
This leads to the overlap of

\begin{align}
\mathrm{F}(t):=\mathrm{Tr}(O^\dagger O(t)) = 1 + it B - \frac{t^2}{2} C + \mathcal{O}(t^3)
\end{align}
where:
\begin{align}
B = \mathrm{Tr}(O^\dagger [H, O]), \qquad 
C = \mathrm{Tr}(O^\dagger [H, [H, O]])
\end{align}
Next we can compute the overlap squarred
\begin{align}
    \left| \mathrm{F}(t)\right|^2 &= \left| 1 + it B - \frac{t^2}{2} C \right|^2 + \mathcal{O}(t^3) \nonumber \\
&= \left(1 + it B - \frac{t^2}{2} C\right)\left(1 - it B^* - \frac{t^2}{2} C^*\right) + \mathcal{O}(t^3) \nonumber \\
&= 1 + 2t \, \mathrm{Im}(B) + t^2 \left( |B|^2 - \mathrm{Re}(C) \right) + \mathcal{O}(t^3)
\end{align}
In our case it holds that $H=H^\dagger$ and $O=O^\dagger$ resulting in $C$ being real. We further make use of $O$ and $H$ being real representations, since they only consists of real pauli-matrices. This results in $B=Tr(O[H,O])$ and $C=\mathrm{Tr}(O[H,[H,O]])$ being real as well, such that the overlap is given by 
\begin{align}
    \abs{\mathrm{F}(t)}^2 &= 1-t^2 (C-B^2) \nonumber \\
    &=1 - t^2 (\mathrm{Tr}(O[H,[H,O]])-\mathrm{Tr}(O[H,O])^2) + \mathcal{O}(t^3).
\end{align}
With this, we can define the zeno-like time 
\begin{align}
    \tau_z^{-2} =\mathrm{Tr}(O[H,[H,O]])-\mathrm{Tr}(O[H,O])^2 
\end{align}
, which results in 
\begin{align}
     \left| \mathrm{F}(t)\right|^2 = 1 -\frac{t^2}{\tau_z^2} + \mathcal{O}(t^3).
\end{align}

If we use the Liouvillian, this timescale is given by
\begin{align}
    \tau_z^{-2} = \mathrm{Tr}(O \mathcal{L}^2) - \left[\mathrm{Tr}(O \mathcal{L})\right]^2
\end{align}
and can be interpreted as an expectation value with respect to the operator $O$, i.e., $\langle \mathcal{L} \rangle = \mathrm{Tr}(O \mathcal{L})$. This leads to
\begin{align}
    \tau_z^{-2} = \langle \mathcal{L}^2 \rangle - \langle \mathcal{L} \rangle^2,
\end{align}
which closely mirrors the quantum Zeno time for state evolution,
\begin{align}
    \tau_{z, (\ket{\psi})}^{-2} = \langle H^2 \rangle - \langle H \rangle^2,
\end{align}
as discussed in \cite{MIS77, FAC08}, justifying the interpretation as a quantum Zeno time for operator evolution.

\bibliography{lit}

\end{document}